\documentclass[12pt,twoside]{article}
\usepackage{fleqn,espcrc1}

\usepackage{graphicx}
\usepackage[figuresright]{rotating}

\title{Antinucleus Production at RHIC}

\author{D. Hardtke\address[LBNL]{Nuclear Science Division, Lawrence Berkeley National Laboratory, Berkeley, CA,  94720} for the STAR Collaboration\footnote{complete author list in J.W. Harris, et al., these proceedings}}%
\begin{document}

\maketitle

\begin{abstract}
Light antinuclei may be formed in relativistic heavy ion collisions via final 
state coalescence of antinucleons.  The yields of antinuclei are sensitive
to primordial antinucleon production, the volume of the system 
at kinetic freeze-out, and space-momentum correlations among antinucleons at 
freeze-out. We report here preliminary
STAR results on $\bar{d}$ and $\overline{^3He}$ production 
in 130A GeV Au+Au collisions.  These results are 
examined in a coalescence framework to elucidate the space-time 
structure of the antinucleon source.  
\end{abstract}

\section{Introduction}

In nucleus-nucleus collisions at RHIC, a nearly baryon free environment with 
high energy density is created in the mid-rapidity region \cite{Calderon,pbarp}.  These conditions 
are favorable for the production of light antinuclei.  
Production of light antinuclei is thought to be possible 
primarily via two mechanisms at RHIC energies.  In the first mechanism, 
nucleus-antinucleus pairs are directly produced in high-energy nucleon-nucleon
or partonic collisions.  This mechanism, however, may not contribute to 
the final observed antinuclei since the small binding energies of the 
produced antinuclei make them extremely susceptible to break-up in the hot
collision zone.  

The second and probably dominant mechanism for production of light antinuclei
is via final state-coalescence.  In this picture, antinucleons of similar
velocity and proximity bind during the late stages of the collision.  In 
this model, the production rates for light antinuclei of antinucleon number $A$ can be related to 
the production rates for constituent 
antinucleons via a coalescence parameter $B_A$ that characterizes the likelihood of antinucleus formation:
\begin{equation}
E \frac{d^3 N_A}{d^3p} = B_A (E \frac{d^3 N_p}{d^3(p/A)})^A.
\label{eqn:BA}
\end{equation}
In the limit where the size of the collision region is smaller than the 
intrinsic size of the produced antinucleus, the coalescence parameter $B_A$
is calculable from the antinuclear wave function \cite{Butler}.  In the limit of large
systems, however, the coalescence parameter can be related to the size of
the system.  In general, the coalescence parameter is shown to scale as $B_A 
\propto (1/V)^{A-1}$ \cite{coalesce} where $V$ is the effective volume where the antinuclei 
are produced.  Coalescence measurements are similar to HBT in that they
give information about the space-time dynamics of the system at freeze-out.

\section{Antinucleus measurements in STAR}

The STAR detector is well-suited for measurement and identification of 
relatively rare antinuclei.  The main tracking detector is a large acceptance ($0<\phi<2\pi,-1.8<\eta<1.8$) Time Projection Chamber(TPC).  We identify antinuclei via the ionization ($dE/dx$) in the TPC.  Figure 1 shows the measured $dE/dx$ plotted against the magnetic rigidity ($p/Z$) for negative tracks.  Also shown are the expectations for $\bar{d}$, $\bar{t}$, and $^3\overline{He}$.  The $\bar{d}$ band is relatively pronounced, and there are 14 counts clustered around the prediction for $^3\overline{He}$.  We observe no evidence of a
$\bar{t}$ band.  Note that we plot here the rigidity, so the mean transverse momentum of the $^3\overline{He}$ sample is 2.4 GeV/c. Under the assumption that $\bar{t}$ and $^3\overline{He}$ are
produced in roughly equal numbers and with similar momentum distributions we
would expect that most $\bar{t}$ would be produced in the high rigidity region
where they cannot be distinguished from more common particles given our $dE/dx$ resolution.    
 
\begin{figure}[htb]
\centering
\mbox{
\includegraphics[width=10cm]{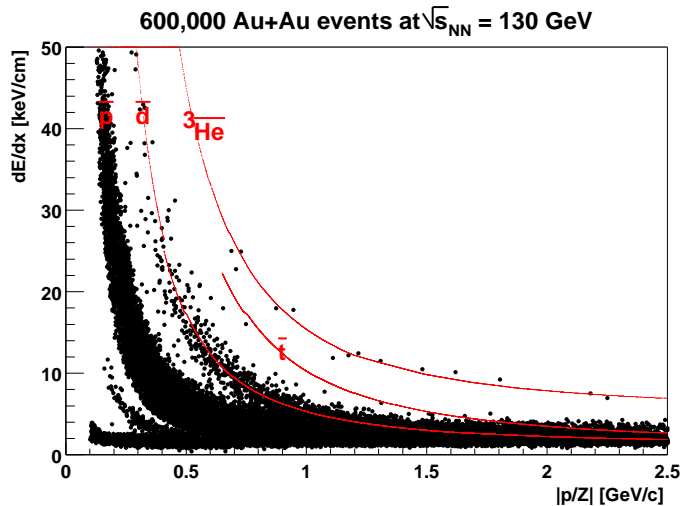}
}
\label{dedx}
\caption{Ionization ($dE/dx$) versus rigidity ($|p/Z|$) for negative tracks.  Also plotted are the Bethe-Bloch expectations for $\bar{d}$, $\bar{t}$ and $^3\overline{He}$.}
\end{figure}

The raw $\bar{d}$ yield is extracted as a function of rapidity ($y$) and transverse
momentum ($p_T$) by fitting the measured $dE/dx$ distribution to an assumed
signal+background form, where the background is from the tail of the $\bar{p}$ $dE/dx$ distribution.  In order to have high $S/B$, we restrict ourselves
to $0.4<p_T<0.8$ GeV/c and $|y|<0.3$.  Once we have obtained a raw yield,
we apply an efficiency correction, 
calculated by embedding simulated tracks into real
events.  We also correct for $\bar{d}$ absorption in the detector, where we
parameterize the $\bar{p}$ absorption in the detector using GEANT simulations
and assume
that $\sigma_{inel.}(\bar{d}) = \sqrt{2}\sigma_{inel.}(\bar{p})$ \cite{Hoang}.  After corrections, we extract $\bar{d}$ invariant yields for the most central 10\% collisions in two $p_T$ bins:
\newpage
\begin{eqnarray*}
\frac{1}{2\pi p_t}\frac{d^2N}{dydp_t} & = & 2.54\pm0.26(stat.)\pm0.64(sys.) \times 10^{-3} GeV^{-2}c^2 \;  [p_T = 0.5 \; \mathrm{GeV/c}] \\ 
 & & 1.89\pm0.19(stat.)\pm0.47(sys.) \times 10^{-3} GeV^{-2}c^2 \; [p_T = 0.7 \; \mathrm{GeV/c}]\\ 
\end{eqnarray*}
Comparing to SPS energies \cite{Bearden}, we see a factor of $\approx 50$ increase in the $\bar{d}$ invariant multiplicity.   

For the $^3\overline{He}$ measurement, statistics are insufficient to do
a differential yield as a function of transverse momentum.  Instead, we use
all 14 counts to extract an invariant yield at the mean $p_T = 2.4$ GeV/c. 
To do this, we calculate a cross-section weighted average efficiency over
the STAR acceptance.  We assume a flat rapidity distribution and an exponential
$p_T$ spectra with $T=0.9$ GeV.  The efficiency is calculated using embedded
tracks in real events, and takes into account the absorption in the detector
assuming  $\sigma_{inel.}(^3\overline{He}) = 2\sigma_{inel.}(\bar{p})$ \cite{Hoang}.  To
increase statistics, we use the top 20\% most central events, and extract an
invariant yield:
 \begin{displaymath}
\frac{1}{2\pi p_t}\frac{d^2N}{dydp_t}  =  8.4\pm2.3(stat.)\pm2.3(sys.) \times 10^{-7} GeV^{-2}c^2 \; [p_T = 2.4 \; \mathrm{GeV/c}]. 
\end{displaymath}
Alternatively, the large $p_T$ acceptance for $^3\overline{He}$ allows us
to extract a dN/dy.  Assuming an exponential $p_T$ distribution and the calculated
efficiency, we minimize a log-likelihood function with the total dN/dy as the
free parameter.  Using this procedure, we extract for the top 20\%
most central events  $\frac{dN}{dy} = 5.1\pm1.7(stat.)\pm1.5(sys.) \times 10^{-5}$.  We have investigated the sensitivity of the extracted dN/dy to the assumed inverse slope parameter, and have found that the final dN/dy differs by
less than 10\% in the range $0.8<T<1.4$ GeV.

\section{Coalescence Comparisons and Discussion}

$\bar{p}$ yields used for comparison and coalescence calculations were presented elsewhere \cite{Harris}. 
We use the measured $\bar{p}$, $\bar{d}$, and $^3\overline{He}$ yields to extract coalescence parameters using equation \ref{eqn:BA}.  The $\bar{p}$ yield
has been corrected for anti-hyperon feed-down using the Hijing event generator
to estimate this contribution to the measured $\bar{p}$ yield.  For $\bar{d}$, we extract $B_2 = 2.13\pm0.20(stat.)\pm0.53(sys.)\times10^{-4} \frac{GeV^2}{c^4}$ over the measured range $0.4<p_T<0.8$ GeV/c and $|y|<0.3$.  Figure 2 shows a comparison with previous measurements.  Comparing to measurements
in similar systems at lower energies we see that the coalescence parameter has dropped by more than a factor of two compared to $\bar{d}$ measurements at SPS
energies.  Assuming $B_2 \propto 1/V$, the antinucleon freeze-out volume increases by $120\pm80\%$ between SPS and RHIC energies.  

Combining the measurements of $\bar{p}$ and $^3\overline{He}$ gives $B_3 = 9.3\pm2.6(stat.)\pm2.8(sys.)\times10^{-8} \frac{GeV^4}{c^6}$.  This is compared
to measurements at lower energies in Figure 3.  NA52 has a 
measurement of $B_3$ for antimatter in nucleus-nucleus, but they observed only
2 counts so the statistics are insufficient for comparison.  Comparing the current results
to the average of the measurements of antimatter and matter at SPS energies,
a noticeable drop in the coalescence parameter is observed.  Assuming $B_3 \propto 1/V^2$, this corresponds to a $180\pm100$\% increase in the antinucleon 
freeze-out volume relative to the SPS.  Within the rather large uncertainties,
the increase in freeze-out volume measured using both $\bar{d}$ and $^3\overline{He}$ is consistent.  

\begin{figure}[htb]
\begin{minipage}[t]{80mm}
\mbox{
\includegraphics[width=8cm]{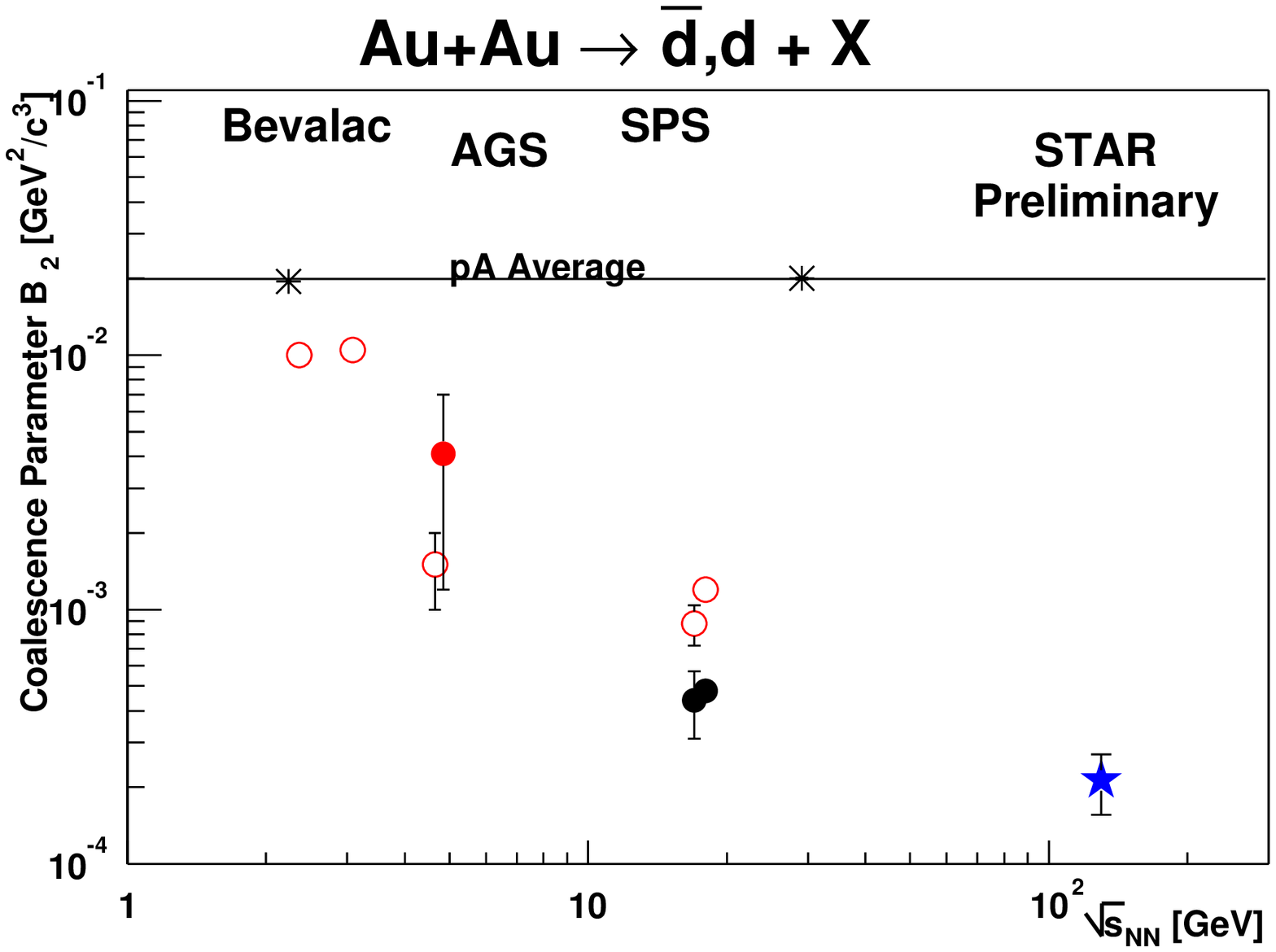}
}
\label{b2ex}
\caption{$B_2$ coalescence parameters versus $\sqrt{s}$ for matter (hollow points) and antimatter (solid points).  Also plotted is the average from $pA$ collisions.}
\end{minipage}
\hspace{0.5cm}
\begin{minipage}[t]{80mm}
\mbox{
\includegraphics[width=8cm]{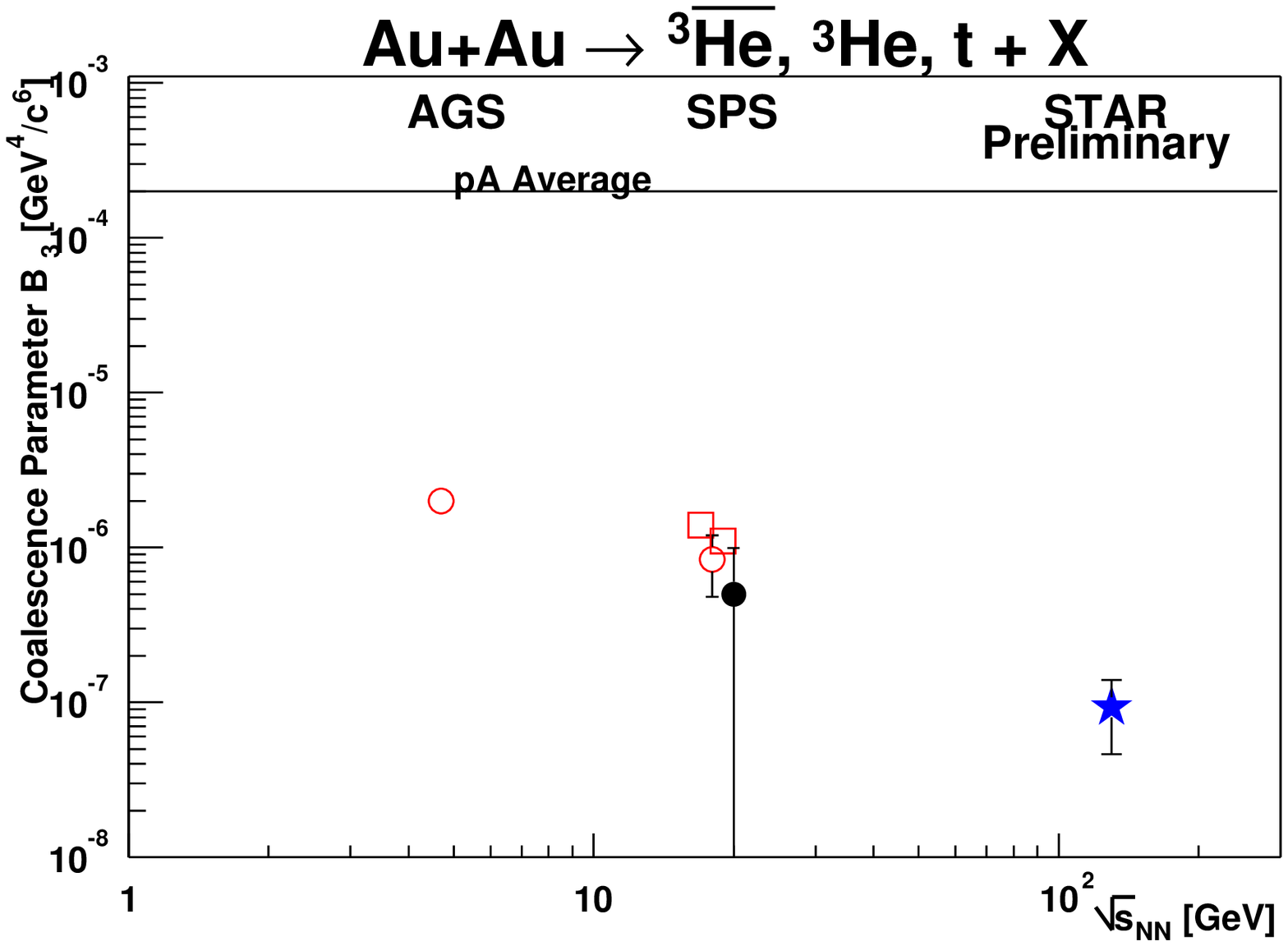}
}
\label{b3ex}
\caption{$B_3$ coalescence parameters versus $\sqrt{s}$ for $\bar{t}$ (hollow squares), $^3He$ (hollow circles) and $^3\overline{He}$ (solid points).  Also plotted is the average from $pA$ collisions.}
\end{minipage}
\end{figure}

\section{Conclusion}

STAR has made the first measurements of light antinuclei at RHIC energies.  
We see a large increase in the antinucleus production rate compared to 
lower energy collisions.
Examined in a coalescence framework, the measured yields indicate more
than a factor of 2 increase in the antinucleon freeze-out volume relative 
to the SPS.

\end{document}